\documentclass[pre,twocolumn,superscriptaddress, aps]{revtex4-2}
\usepackage[utf8]{inputenc}

\usepackage[separate-uncertainty = true,multi-part-units=single]{siunitx}
\usepackage{changes}
\usepackage{graphicx}
\usepackage{amsmath}
\usepackage{amssymb}
\usepackage{todonotes}

\usepackage{mhchem}
\usepackage{chemformula}
\usepackage{algorithm} %
\usepackage[noend]{algpseudocode} 

\usepackage[shortlabels]{enumitem} 

\usepackage[english]{babel}
\usepackage[hidelinks]{hyperref}

\setchemformula{kroeger-vink}
\DeclareSIUnit\angstrom{\text{Å}}

\presetkeys%
    {todonotes}%
    {inline,backgroundcolor=white}{}
\makeatletter

\makeatletter
\renewcommand{\todo}[2][]{\@bsphack\@todo[#1]{\textcolor{black!70}{#2}}\@esphack\ignorespaces}

\makeatother

\begin{document}
\title{Thermodynamics of alkali feldspar solid solutions with varying Al–Si order: atomistic simulations using a neural network potential}

\author{Alexander Gorfer}
\affiliation{Faculty of Physics, University of Vienna, Boltzmanngasse 5, 1090, Vienna, Austria}
\affiliation{Department of Lithospheric Research, University of Vienna, Josef-Holaubek-Platz 2, 1090, Vienna, Austria}
\author{David Heuser}
\affiliation{Department of Lithospheric Research, University of Vienna, Josef-Holaubek-Platz 2, 1090, Vienna, Austria}
\author{Rainer Abart}
\affiliation{Department of Lithospheric Research, University of Vienna, Josef-Holaubek-Platz 2, 1090, Vienna, Austria}
\author{Christoph Dellago}
\affiliation{Faculty of Physics, University of Vienna, Boltzmanngasse 5, 1090, Vienna, Austria}

\begin{abstract}
The thermodynamic mixing properties of alkali feldspar solid solutions between the Na and K end members were computed through atomistic simulations using a neural network potential. We performed combined molecular dynamics and Monte Carlo simulations in the semi-grand canonical ensemble at \SI{800}{\celsius} and considered three quenched disorder states in the Al\nobreakdash-Si\nobreakdash-O framework ranging from fully ordered to fully disordered. The excess Gibbs energy of mixing, excess enthalpy of mixing and excess entropy of mixing are in good agreement with literature data. In particular, the notion that increasing disorder in the Al\nobreakdash-Si\nobreakdash-O framework correlates with increasing ideality of Na-K mixing is successfully predicted. Finally, a recently proposed short range ordering of Na and K in the alkali sublattice is observed, which may be considered as a precursor to exsolution lamellae, a characteristic phenomenon in alkali feldspar of intermediate composition leading to perthite formation during cooling.
\end{abstract}
\maketitle

\section{Introduction}\label{sec:intro}
\begin{figure}[t]
    \centering
    \makebox[\columnwidth][c]{\includegraphics[width=1.0\columnwidth]{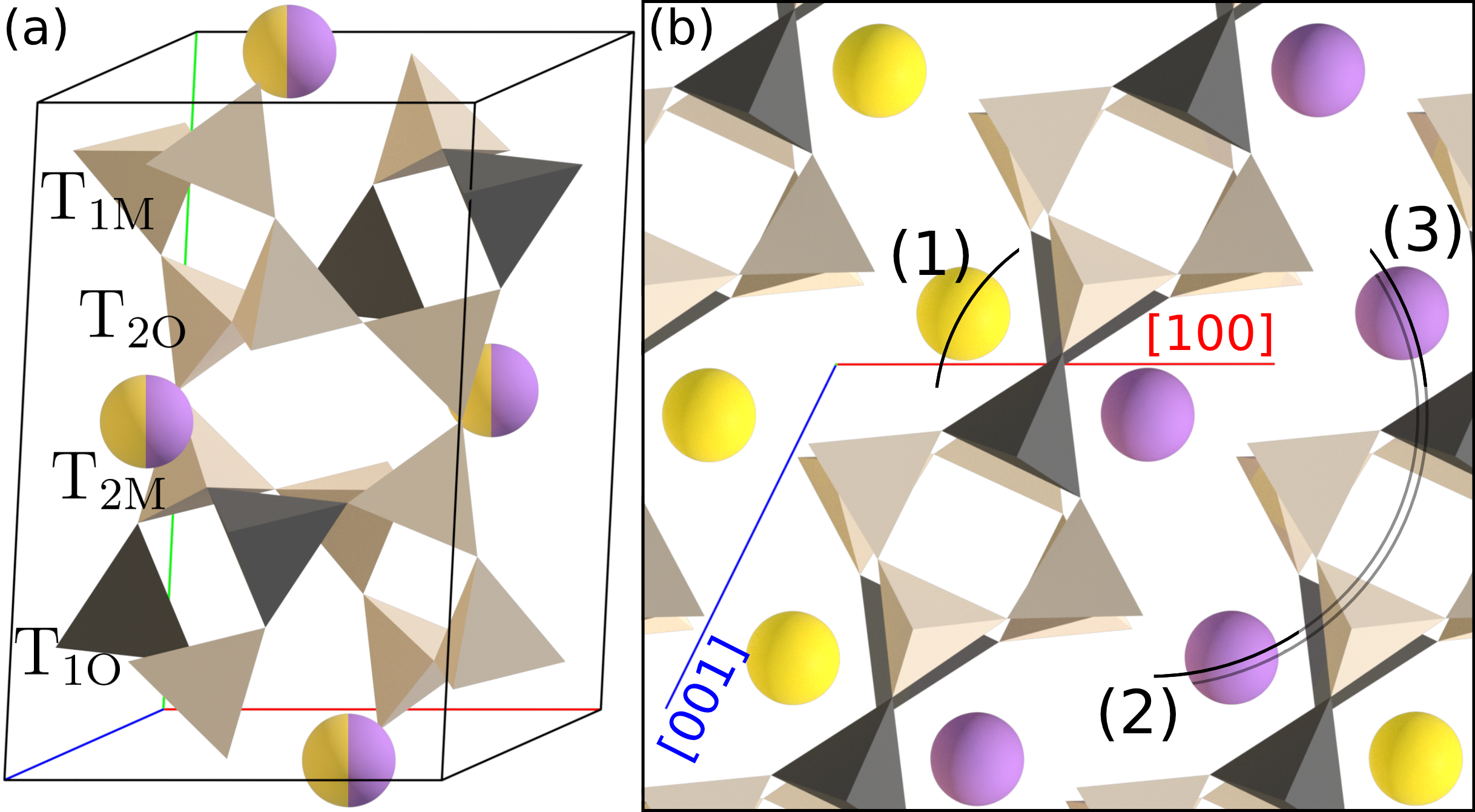
    }}
    \caption{(\(\mathbf{a}\)) Unit cell of \ch{Al} ordered alkali feldspar with \ch{Al^{3+}} at the \(\mathrm{T}_{\mathrm{1O}}\) tetrahedra. The Al-Si tetrahedra types are labeled on the leftmost tetrahedra. \ch{Na^+}/\ch{K^+} is illustrated as yellow/purple spheres. In (\(\mathbf{b}\)) we see the \((010)\) plane and the first three Na-K configuration shells. The \ch{Na^+}/\ch{K^+} distribution shows the SRO we found in an exaggerated fashion.
    }\label{fig:uni_cell}
\end{figure}

Feldspar, the most abundant mineral in the Earth's crust, is a framework silicate comprised of corner sharing \ce{[SiO4]^{-4}} and \ce{[AlO4]^{-5}} tetrahedra forming a three-dimensional framework with large cavities, which are occupied by cations that balance the net negative charge of the tetrahedral framework. The most common cations with suitable ionic radius and charge to enter these cavities are Ca$^{2+}$, Na$^+$ and K$^+$, giving raise to the ternary solid solution: \ch{CaAl2Si2O8}-\ch{NaAlSi3O8}-\ch{KAlSi3O8}. The binary solid solution between \ch{NaAlSi3O8} (albite) and \ch{KAlSi3O8} (K-feldspar) is known as alkali feldspar, which is further classified based on the potassium site fraction on the alkali site \(X_{\ch{K}}\) and on the degree of disorder in the Al\nobreakdash-Si\nobreakdash-O tetrahedral framework. Different states of disorder may arise due to the fact that there are four distinct tetrahedral sites: \(\mathrm{T}_{\mathrm{1O}}, \mathrm{T}_{\mathrm{1M}}, \mathrm{T}_{\mathrm{2O}}\) and \(\mathrm{T}_{\mathrm{2M}}\) (Fig.~\ref{fig:uni_cell}~a). In low-temperature, ordered alkali feldspar, the \ch{Al^{3+}} cations exclusively occupy the \(\mathrm{T}_{\mathrm{1O}}\) sites. At intermediate temperatures, \ch{Al^{3+}} favors \(\mathrm{T}_{\mathrm{1O}}\) and \(\mathrm{T}_{\mathrm{1M}}\) over \(\mathrm{T}_{\mathrm{2O}}\) and \(\mathrm{T}_{\mathrm{2M}}\), and at high temperatures \ch{Al^{3+}} is randomly distributed over all the tetrahedral sites.

The disorder phenomena that accompany Na\nobreakdash-K mixing on the alkali sublattice and Al\nobreakdash-Si mixing on the Al\nobreakdash-Si\nobreakdash-O framework make alkali feldspar a challenging system in terms of its thermodynamic properties.
The Al\nobreakdash-Si disorder in particular can take significant timescales to equilibrate, like the fully ordered low microcline in Klokken that took \num{e4} to \num{e5} years to develop \cite{brown_exsolution_1984}. This means that many laboratory experiments %
are conducted on essentially quenched states of Al-Si disorder.
Therefore, investigations into feldspar thermodynamics must consider \(T\), \(P\), \(X_{\ch{K}}\) and the state of Al\nobreakdash-Si disorder. The latter is typically classified using Kroll and Ribbe's scheme~\cite{kroll_determining_1987}, which is based on unit cell dimensions. An accurate understanding of feldspar thermodynamics is key in deducing the thermal history of magmatic and metamorphic rocks from the compositions and microstructures of alkali feldspars in exhumed samples.

A recent overview of the current state of knowledge about alkali feldspar thermodynamics can be found in Ref.~\cite{hovis_refined_2017}. Over the years, several authors have performed experiments on the thermodynamics of alkali feldspar. The solvus, which has its \(T_\mathrm{C}\) around 650°C to 870°C, depending on the state of Al-Si disorder and on pressure, was directly determined for various states of  Al-Si disorder primarily in the 1960s and 70s but also in 2024 \cite{orville_alkali_1963, luth_alkali_1966, bachinski_experimental_1971, smith_alkali-feldspar_1974, goldsmith_experimental_1974, lagache_system_1977, parsons_alkali-feldspars_1978, heuser_coherent_2024}. Sodium-potassium partitioning experiments have been performed repeatedly, starting from 1975 and up to 2024 \cite{delbove_excess_1975, hovis_gibbs_1991, neusser_experimental_2012, heuser_thermodynamic_2024}. Also measured were mixing enthalpies for different ordering states in \cite{hovis_gibbs_1991, hovis_refined_2017} and vibrational entropies in \cite{haselton_calorimetric_1983, heuser_thermodynamic_2024}.

The Al-Si disorder also complicates computational studies, as modeling disorder generally requires large system sizes due to the lack of translational symmetry.  %
This calls for application of computationally inexpensive classical force-fields, which are tuned empirically, such as lattice energy models that have 
been applied to Al\nobreakdash-Si thermodynamics in feldspar \cite{zhang_si_2007, dubacq_thermodynamics_2022}. It is, however, desirable to use first-principles methods such as Density Functional Theory (DFT), which can provide higher accuracy with less reliance on human tuning. But due to their computational costs, studies using DFT have either focused on properties of ordered end-members \cite{kaercher_ab_2014, benisek_accuracy_2018,antonelli_kinetic_2019, li_first-principles_2019-1}, or used systems with essentially "hand-made" Na-K disorder with limited realizations of \(X_{\ch{K}}\) to study the vibrational entropy~\cite{benisek_first-principles_2015}, excess enthalpy of mixing~\cite{benisek_excess_2020} and K-O bond lengths \cite{li_first-principles_2019}. Other features of alkali feldspar that were obtained by DFT modeling include point defects~\cite{gorfer_structure_2024} and the feldspar-water interface~\cite{pedevilla_can_2016, dickbreder_atomic_2024, piaggi_first-principles_2024}.

To summarize, despite significant effort, the thermodynamics of alkali feldspar remain incompletely understood. Machine Learning Force Fields (MLFF), reviewed in \cite{unke_machine_2021}, offer a promising pathway towards a better understanding by combining the accuracy of DFT calculations with system scales accessible only to classical force-field methods. Recent studies have already explored the applicability of MLFFs for alkali feldspar, including a neural network potential for \ch{Na^+} point defects in albite~\cite{gorfer_structure_2024} to study Na-diffusion~\cite{gorfer_mechanism_2024} and, motivated by the ice-nucleating properties of alkali-feldspar, a deep potential for modeling the microcline-water interface \cite{piaggi_first-principles_2024}.

This study investigates the thermodynamic properties of the alkali feldspar solid solution across the full range of Na-K compositions and formally treats disorder for three archetypes:
\begin{enumerate}[nosep]
    \item \textit{Al ordered} with \(X_{\mathrm{T1O}} = 1\). This ordering is supposed to represent solid solutions between low albite and microcline.
    \item \textit{Al \(\mathit{T}_{\mathit{1}}\)-disordered} with \(X_{\mathrm{T1O}} = X_{\mathrm{T1M}} = 0.5\). This ordering is supposed to represent an in-between configuration with \ch{Al} occupying both \(\mathrm{T}_{1\mathrm{O}}\) and \(\mathrm{T}_{1\mathrm{M}}\) at random such as may be found in orthoclase.
    \item \textit{Al disordered} with \(X_{\mathrm{Ti}} = 0.25\). This ordering represents the most extreme type of disorder, such as may be found in solid solutions between high albite and sanidine.
\end{enumerate}

We discuss an algorithm for creating specific states of Al\nobreakdash-Si and Na\nobreakdash-K disorder, the training of a Neural Network Potential (NNP) to simulate their dynamics, and Semi-Grand Canonical Monte Carlo (SGCMC) simulations to calculate equilibrium \(X_{\ch{K}}\) as a function of chemical potential. The results, which include the excess Gibbs energy of mixing, excess enthalpy of mixing and excess entropy of mixing  show good agreement with experimental data, and they also reveal the impact of Al-Si disorder on these excess properties as well as on the solvus. In addition, we observe and quantify a Short Range Ordering (SRO) phenomenon leading to a negative excess configurational entropy and classify it as a precursor to exsolution, which typically occurs in alkali feldspar of intermediate composition during cooling and leads to characteristic lamellar intergrwoth of Na-rich and K-rich alkali feldspar referred to as {\em perthite}.

\section{Methods}

\subsection{Na-K short range order}
To quantify the degree of Na\nobreakdash-K disorder within the alkali sublattice, we use Warren-Cowley Short Range Order (SRO) parameters
\begin{equation}\label{eq:SRO}
    \alpha_{\ch{NaK}}^{(l)} = 1 - \frac{Z_{\ch{K}}^{(l)}}{Z_{\mathrm{tot}}^{(l)} X_{\ch{K}}}.
\end{equation}
These are calculated for every \ch{Na} ion, where \(Z_{\ch{K}}^{(l)}\) is the instantaneous number of \ch{K} ions within the \(l\)-th configuration shell, %
\(Z_{\mathrm{tot}}^{(l)}\) is the total number of neighbors in the \(l\)-th shell, and \(X_{\ch{K}}\) is the overall \ch{K} site fraction of K on the alkali sublattice. In a completely random alloy, the expected number of \ch{K} neighbors for an \ch{Na} would be \(Z_{\mathrm{tot}}^{(l)} X_{\ch{K}}\) and \(\alpha_{\ch{Na}}^{(l)} = 0\). If \(\alpha_{\ch{Na}}^{(l)} > 0\), there is a tendency for \ch{Na} and \ch{K} ions to cluster, while for \(\alpha_{\ch{Na}}^{(l)} < 0\) there exists an ordering where \ch{Na} ions have more \ch{K} neighbors than expected by chance. In Fig.~\ref{fig:uni_cell}~(b) the first three Na-K configuration shells are illustrated. Due to their similarities, for the second and third nearest neighbors we calculate an \(\alpha_{\ch{Na}}^{(2-3)}\) as if they were on the same shell. 

\subsection{Treating Al-Si and Na-K disorder in alkali feldspar}\label{sec:disorder_def}

As discussed in the Introduction, the state of Al\nobreakdash-Si disorder in the  tetrahedral framework may be a complicated function of temperature, pressure as well as potassium site fraction, but the slow ordering of the Al\nobreakdash-Si~\cite{grove_coupled_1984, cherniak_silicon_2003} allows, or more appropriate, forces experiments to be done with a quenched Al\nobreakdash-Si order on laboratory timescales. While this is less true at and above certain temperatures and timescales where Al-Si disordering is reported~\cite{hovis_behavior_1986, heuser_thermodynamic_2024}, since our aim is to best emulate usual experimental conditions, it is reasonable to assume a state of quenched Al\nobreakdash-Si disorder that does not change throughout the simulation.

Atomistic simulations of disordered systems require large system sizes as the translational symmetry is broken. For first-principle electronic structure calculations with DFT, where computational cost scales with the third power of the number of electrons, these system sizes are prohibitively expensive and have only recently been investigated using specific pre-defined configurations of Na\nobreakdash-K disorder by Benisek and Dachs~\cite{benisek_first-principles_2015, benisek_excess_2020} and by Li et al.~\cite{li_first-principles_2019}. %

We propose an algorithm to generate well-defined configurations of disordered alkali feldspars. Below is its pseudocode, but we also give an explanation in prose. 

The algorithm takes as input the potassium fraction \(X_{\ch{K}}\), the \ch{Al}-fractions at each of the different tetrahedral sites \(\{X_{\mathrm{T1O}}, X_{\mathrm{T1M}}, X_{\mathrm{T2O}}, X_{\mathrm{T2M}} \}\), the supercell dimension in each direction \(a, b, c\) and a parameter that controls the Na\nobreakdash-K distribution scheme. 

\begin{algorithm}[H]
    \renewcommand\thealgorithm{}
    \caption{Alkali feldspar disorder generation scheme}
    \label{alg_disgen}
    \begin{algorithmic}[1]
        \Require{Fractions \(X_{\ch{K}}\), \(X_{\mathrm{T1O}}\), \(X_{\mathrm{T1M}}\), \(X_{\mathrm{T2O}}\), \(X_{\mathrm{T2M}}\) and 
        
        integer supercell dimensions \(a, b, c\) and
        
        bool NaK-SQS}
        \State Create supercell configuration with dimensions \(a, b, c\).
        \State Distribute Al among possible tetrahedra at random.
        \State Swap Al with Al neighbors with Si without any Al neighbors to fulfill Loewenstein’s rule.
        \While{\(X_{\mathrm{T}i}' \neq X_{\mathrm{T}i}\)}
        \State move Al from site with \(X_{\mathrm{T}i}' > X_{\mathrm{T}i}\) to site with \(X_{\mathrm{T}i}'~<~X_{\mathrm{T}i}\).
        \EndWhile
        \If{NaK-SQS}
            \State Distribute Na-K according to SQS.
        \Else
            \State Distribute Na-K at random.
        \EndIf
        \State \Return{configuration}
    \end{algorithmic}
\end{algorithm}

To create Al\nobreakdash-Si disorder, the aluminum is distributed randomly across all Al\nobreakdash-Si tetrahedra with non-zero \ch{Al}-fraction. Adherence to Loewenstein's rule, which prohibits adjacent \ch{Al} tetrahedra and is assumed to be well fulfilled in alkali feldspar, is maintained by iteratively swapping an \ch{Al} that neighbors an \ch{Al} by an \ch{Si} that is surrounded exclusively by \ch{Si}. The algorithm then compares the instantaneous \ch{Al}-fractions (\(X_{\mathrm{T}_{i}}'\)) with the input values and continuously swaps \ch{Al} and \ch{Si}, adhering to Loewenstein's rule, until the desired \ch{Al}-fractions are reached. At \textit{unreasonable} input \ch{Al}-fractions such as \(X_{\mathrm{T}_{\mathrm{1M}}} = X_{\mathrm{T}_{\mathrm{2M}}} = 0.5,~ X_{\mathrm{T}_{\mathrm{1O}}} = X_{\mathrm{T}_{\mathrm{2O}}} = 0\) we observed that it is possible for the algorithm to get stuck in a configuration and allow a random move to get out, but for realistic Al-disorder states and in particular for all disorder states treated in this work, this is not necessary.

To generate a disordered alkali-sublattice we either just distribute Na\nobreakdash-K at random, or we distribute them according to Special Quasirandom Structures (SQS). In the latter scheme, the Na\nobreakdash-K configuration is optimized to get \(\alpha_{\ch{Na}}^{(l)}\) as close to zero as possible. Using SQS speeds up the convergence of self-averaging properties with increasing system-sizes. Our Python implementation, openly available on GitHub, employs the sqsgenerator package~\cite{gehringer_models_2023} to generate SQS and optimizes using the first 10 shells. It also uses the ASE package~\cite{larsen_atomic_2017} to represent atomic configurations. %

In this study we consider the three types of archetypical Al\nobreakdash-Si disorder that are defined in the Introduction: Al ordered, Al \(\mathrm{T}_1\)-disordered and Al disordered.

The literature offers algorithms in which the Al-disorder configuration results from simulations using empirical lattice energy models \cite{zhang_si_2007, dubacq_thermodynamics_2022}. While the result of these modeling approaches can be expected to offer higher realism in the Al\nobreakdash-Si disorder configurations, in this study we wanted to rely on a purely first-principles methodology such that we refrained from using empirical force-fields to create the Al-Si ordering and we also wanted to create a baseline of extreme ordering archetypes that future investigations may build upon.

\subsection{Ab initio calculations}
To train and test the NNP, with which we ultimately run large-scale (\num{19968} atoms) simulations, we performed smaller-scale (\(\sim\)416 atoms) DFT calculations using the Vienna Ab initio Simulation Package (VASP) \cite{Kresse_VASP_1_1993,Kresse_VASP_2_1996,Kresse_VASP_3_1996} with the projector augmented wave method \cite{KressePAW1999} using the Perdew-Burke-Ernzerhof (PBE) \cite{perdew_generalized_1996} exchange correlation functional. We used identical settings and convergence criteria as in Ref.~\cite{gorfer_structure_2024}, allowing us to reuse the associated dataset of that study~\cite{gorfer_structure_2024-1}. When charged point defects are included in the supercell, we varied the number of electrons using a jellium background and included the point-charge electrostatic correction to the potential energy to correct for finite-size electrostatic effects (see SM for details). 

\subsection{Neural network potential and training/testing datasets}
We developed a Neural Network Potential (NNP) that can treat Al\nobreakdash-Si disorder, mixtures of Na\nobreakdash-K as well as the following point-defects: Na-Na-dumbbell, \ch{Na}-\((0,0,\frac{1}{2})\)-interstitial, Na-K-dumbbell, Na-vacancy and K-vacancy (see \cite{gorfer_structure_2024} for nomenclature), although the point defects are immaterial to this study. Our NNP uses the architecture of Behler and Parrinello~\cite{behler_generalized_2007}.

The local atomic environments are represented by radial symmetry functions and two classes of angular atom-centered symmetry functions. For angles involving \ch{O}, we use both the narrow and wide angular symmetry functions given in Ref.~\cite{behler_atom-centered_2011}. To also describe all other possibilities, we use weighted atom-centered angular symmetry functions (wACSF) \cite{gastegger_wacsfweighted_2018}. Unlike \cite{gastegger_wacsfweighted_2018}, to discern between different chemical elements, our wACSFs use the Pauling electronegativity instead of the atomic number as variables in the element-dependent weighting function \(h\) (see the SM for details). The input layer of the feed forward networks, consisting of the symmetry functions, has a length of 213 for the case of \ch{O}, 209 for \ch{Al}, 210 for \ch{Si}, 211 for \ch{Na} and 205 for \ch{K}. This is followed by two hidden layers with 25 nodes each and a single output node. We trained four potentials with random initial weights using the n2p2 package~\cite{singraber_library-based_2019}, optimizing the root mean squared error via a parallel Kalman-filter~\cite{singraber_parallel_2019} and combine them in a committee machine using the n2p2-committee implementation given in Ref.~\cite{kyvala_diffusion_2023}.

To construct the training and testing datasets we first reused the training set constructed in Ref.~\cite{gorfer_structure_2024} for Na-feldspar, except that we apply the point-charge electrostatic correction instead of the correction of Kumagai and Oba (see SM). We then also employed the on-the-fly approach described in \cite{gorfer_structure_2024} using the on-the-fly learning implementation of VASP~\cite{jinnouchi_--fly_2019, jinnouchi_phase_2019} for a hand-made Na-K mixed \(2\!\times\!1\!\times\!2\) supercell containing 13 \ch{K} and 3 \ch{Na}, both with and without an \ch{Na^+}-interstitial. The combined dataset was used to create a preliminary NNP committee, which we iteratively refined through an active-learning scheme similar to Ref.~\cite{gorfer_structure_2024} by incorporating uncertain configurations encountered during MD simulations. The differences from Ref.~\cite{gorfer_structure_2024} are the supercell sizes, which are \(2\!\times\!1\!\times\!2\), the variation of the Na\nobreakdash-K distribution covering \(X_{\ch{K}}\) from 0.0 to 1.0, the use of hand-made Al\nobreakdash-Si disorder realizations and the use of the aforementioned point-defects. After repeating this 10 times, creating different disorder realizations in each iteration, we had a training set of 4657 structures as well as a hyperparameter-optimized and stable force-field. We then used our algorithm to further expand the dataset using systematically generated structures: \(2\!\times\!2\!\times\!2\) supercells with Al disorder were generated for potassium fractions \(X_{\ch{K}} = \{0.1, 0.2, \dots 0.9\} \), where for each \(X_{\ch{K}}'\), 10 different Al-Si and Na-K disorder realizations were generated (90 defect-free realizations in total). The NaK\nobreakdash-SQS scheme was omitted, as we wanted to cover all possible local motifs. In these systems, the 5 defect-types were introduced separately (540 initial systems). These were fully relaxed to create \SI{0}{\kelvin} configurations. Subsequently, for each \(X_{\ch{K}}'\), 6 realizations including the defect-free and the defective systems were simulated using NPT-MD at \SI{600}{\kelvin}, then \SI{1000}{\kelvin} and continuing at \SI{1400}{\kelvin}. Each temperature was simulated for \SI{60}{\pico\second} and configurations at \(20, 25, \dots 60\) \si{\pico\second} were saved (regardless of committee uncertainty). Configurations from 2 realizations per \(X_{\ch{K}}'\), which were simulated using NPT-MD, were used to build a test set containing 1667 structures, while the rest were added to the training set which includes 8720 structures in total. %

\subsection{Neural network potential performance}
\begin{figure}
    \centering
    \makebox[\columnwidth][c]{\includegraphics[width=1\columnwidth]{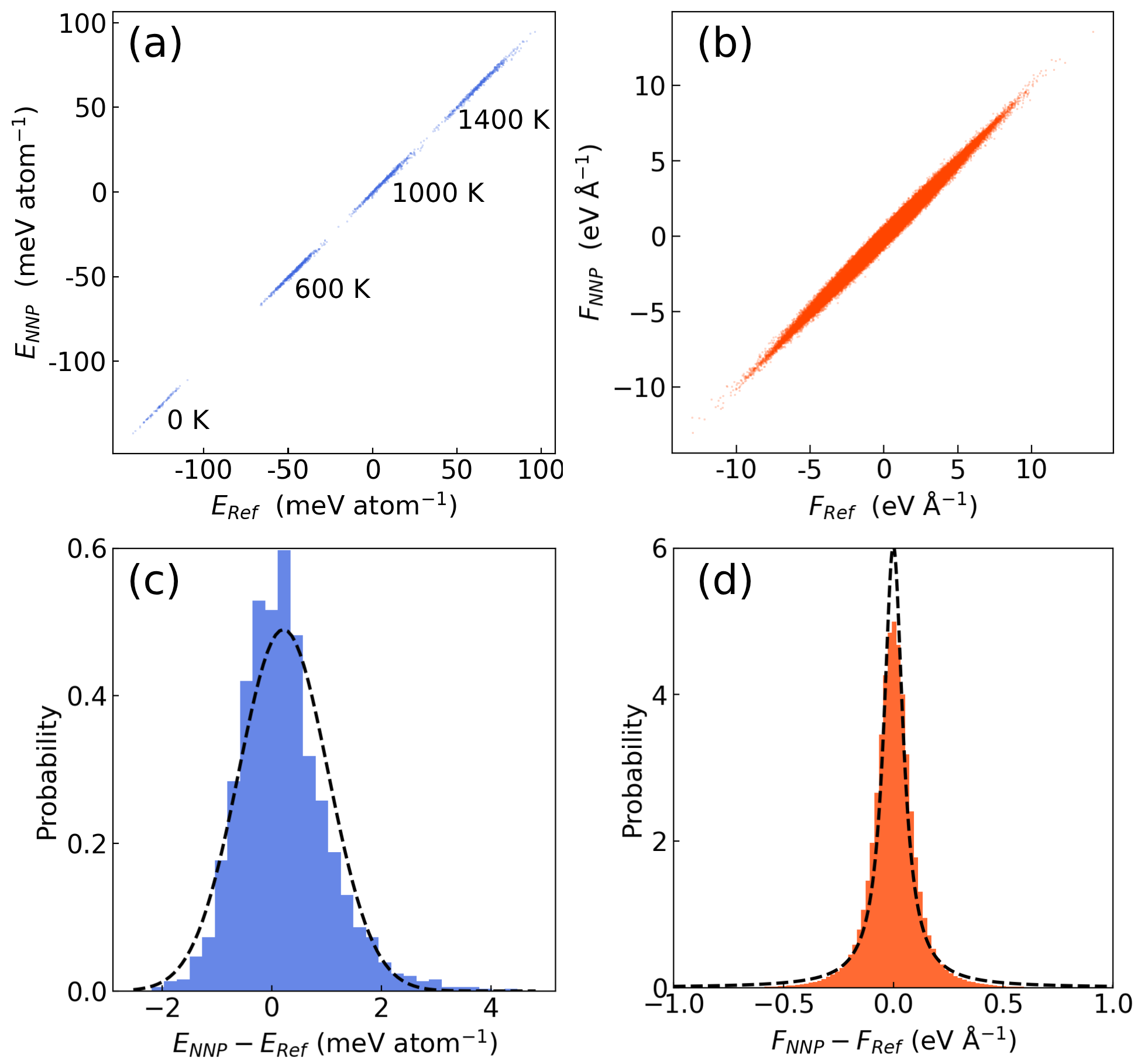
    }}
    \caption{(\(\mathbf{a}\)) Energy parity plot for the test set showing the neural network potential (NNP) versus the DFT reference. The axes were centered at zero.  The four different temperatures  sampled in the test set can be seen. (\(\mathbf{b}\))  parity plot for the force predictions. The error distributions are shown in the second row. For energy errors (\(\mathbf{c}\)) a normal distribution is overlain, for force errors (\(\mathbf{d}\)) a Cauchy distribution is shown instead.
    }   \label{fig:test_results}
\end{figure}

The NNP was tested on a set containing 1667 Al disordered structures covering \(X_{\ch{K}} = \{0.1, 0.2, \dots 0.9\} \) and the defects: Na-Na-dumbbell, \ch{Na}-\((0,0,\frac{1}{2})\)-interstitial, Na-K-dumbbell, Na-vacancy and K-vacancy as well as the defect-free system for temperatures 0, 600, 1000 and \SI{1400}{\kelvin}. The specific realizations of Na\nobreakdash-K and Al\nobreakdash-Si disorder states that are tested are not included in the training set to test the ability to generalize to new unseen configurations. The potential performs well over the whole dataset as can be seen from the root mean squared errors (RMSE) and mean average errors (MAE) shown in Tab.~\ref{Tab:rmse} and errors for each configuration separately in Fig.~\ref{fig:test_results}.

\begin{table}[!h]
    \centering
    \caption{Energy and force RMSE and MAE for the test set.}
    \begin{tabular}{l | c c }
           & \(\Delta E\) \(\left[\tfrac{\text{meV}}{\text{atom}}\right]\) & \(\Delta F\) \(\left[\tfrac{\text{eV}}{\si{\angstrom}}\right]\) \\\toprule
      RMSE & 0.84   & 0.12 \\
      MAE  & 0.63   & 0.08 \\\hline
    \end{tabular}
    \label{Tab:rmse}
\end{table}

\subsection{Unit cell minimization}
To determine the equilibrium \SI{0}{\kelvin} unit cell parameters we used our algorithm to generate supercells with dimensions \(8\!\times\!6\!\times\!8\) (i.e.~19968 atoms) for the three Al\nobreakdash-Si disorder states covering \(X_{\ch{K}} = \{0, 0.1, \dots 1\}\) with NaK\nobreakdash-SQS enabled. Each system was then fully relaxed in LAMMPS~\cite{thompson_lammps_2022} by iterating between purely ionic relaxations and relaxations of both ionic and box degrees of freedom (DOF). Both minimizations consisted of a conjugated gradient descent followed by a steepest gradient descent (SGD). The purely ionic SGD always reached the energy threshold of \SI{1e-12}{\electronvolt}, whereas initially, the ionic plus box DOF SGD had to be preemptively terminated after a maximum of 3000 steps was reached.
Iterations between purely ionic and ionic plus box DOF relaxations were repeated until %
the energy difference between the final configuration and the next-to-last box-DOF SGD relaxed structure was less than \SI{0.01}{\electronvolt} or \SI{6.3 e-4}{\kilo\joule\per\mole} for all structures. The disordered systems with \(X_{\ch{K}} = 0.0\) required the longest with around 30 iterations to reach this convergence level. Input files are available in the Zenodo repository~\cite{gorfer_thermodynamics_2024}.

\subsection{Miscibility calculations using SGCMC-MD}
To determine mixing thermodynamics at finite temperature we simulated alkali feldspar using Molecular Dynamics (MD) together with Semi-Grand Canonical %
Monte Carlo (SGCMC), implemented within LAMMPS~\cite{thompson_lammps_2022, sadigh_scalable_2012}. The SGCMC approach allows alchemical transformations from \ch{Na} to \ch{K} as well as \ch{K} to \ch{Na} that are accepted with a probability \cite{sadigh_scalable_2012}:
\begin{equation*}
    P_{\mathrm{acc}} = \min \{1, \exp \left[-\beta (\Delta U + \Delta \mu N \Delta X_{\ch{K}})\right]\},
\end{equation*}
where \(\Delta U\) is the change in potential energy before and after alchemical swapping, \(\Delta X_{\ch{K}}\) is the change in potassium site fraction and \(\Delta \mu \equiv \mu_{\ch{Na}} - \mu_{\ch{K}}\) %
is the difference in chemical potential between \ch{K} and \ch{Na} which is set before every simulation. Aiming to cover the whole range of Na-K compositions, we identified the following suitable values of \(\Delta \mu = \) \{\num{-1}, \num{-0.05},  \num{-0.0375},  \num{-0.03125},  \num{-0.028125},  \num{-0.0265625},  \num{-0.025},  \num{-0.021875}, \num{-0.01875},  \num{-0.0125},  \num{-0.00625},  \num{0}, \num{0.0125}\}   (\si{\electronvolt}) by trial and error.

For each Al-ordering state and \(\Delta \mu\) value, a relaxed \(8\!\times\!6\!\times\!8\) supercell was first equilibrated in the SGC-PT ensemble at \SI{1073.15}{\kelvin} and atmospheric pressure. A single realization of Al-ordering was used for each of the three Al-ordering archetypes. The damping parameters were \SI{0.1}{\pico\second} for the Langevin thermostat and \SI{1}{\pico\second} for the Nose-Hoover barostat. The timestep of the dynamics was \SI{1}{\femto\second}. Initially, 5000 MC moves were attempted every 400 timesteps. After \num{4 e5} MC attempts, the frequency was reduced to 250 attempts per 400 timesteps and the equilibration was continued for an additional \num{5 e5} attempts. Subsequently, while maintaining these parameters, configurations including the chemical identity for every atom, the potential energy and volume was written to disk every 400 timesteps at exactly one timestep before the MC moves. These production runs were conducted for an additional \num{7 e5} MC attempts, lasting around \SI{1.1}{\nano\second}. The enthalpy of the end-members was calculated in the NPT ensemble without SGCMC using the aforementioned thermo- and barostat.  %

The acceptance ratio depends on \(\Delta \mu\). While it was essentially zero at \SI{-1}{\electronvolt}, at \SI{-0.021875}{\electronvolt} it was \(27\%\). %
We note that we also attempted MC moves involving swaps between Al and Si, but no moves were ever accepted even for a trial run at \SI{1400}{\kelvin}, implying that the local environment of Al and Si is too different for a sufficient acceptance probability. This is unfortunate as a working SGCMC simulation methodology for the Al\nobreakdash-Si sublattice would allow us to calculate the equilibrium Al\nobreakdash-Si disorder states, in similarity to Dubacq~\cite{dubacq_thermodynamics_2022}.

\section{Results and Discussion}

\subsection{Crystal symmetry}
We used fully relaxed supercells (dimensions \(8\!\times\!6\!\times\!8\), 19968 atoms) to determine the unit cell parameters across the entire range of potassium site fractions \(X_{\ch{K}}\) for the three ordering states defined in the Introduction.

\begin{figure}
    \centering
    \makebox[\columnwidth][c]{\includegraphics[width=1\columnwidth]{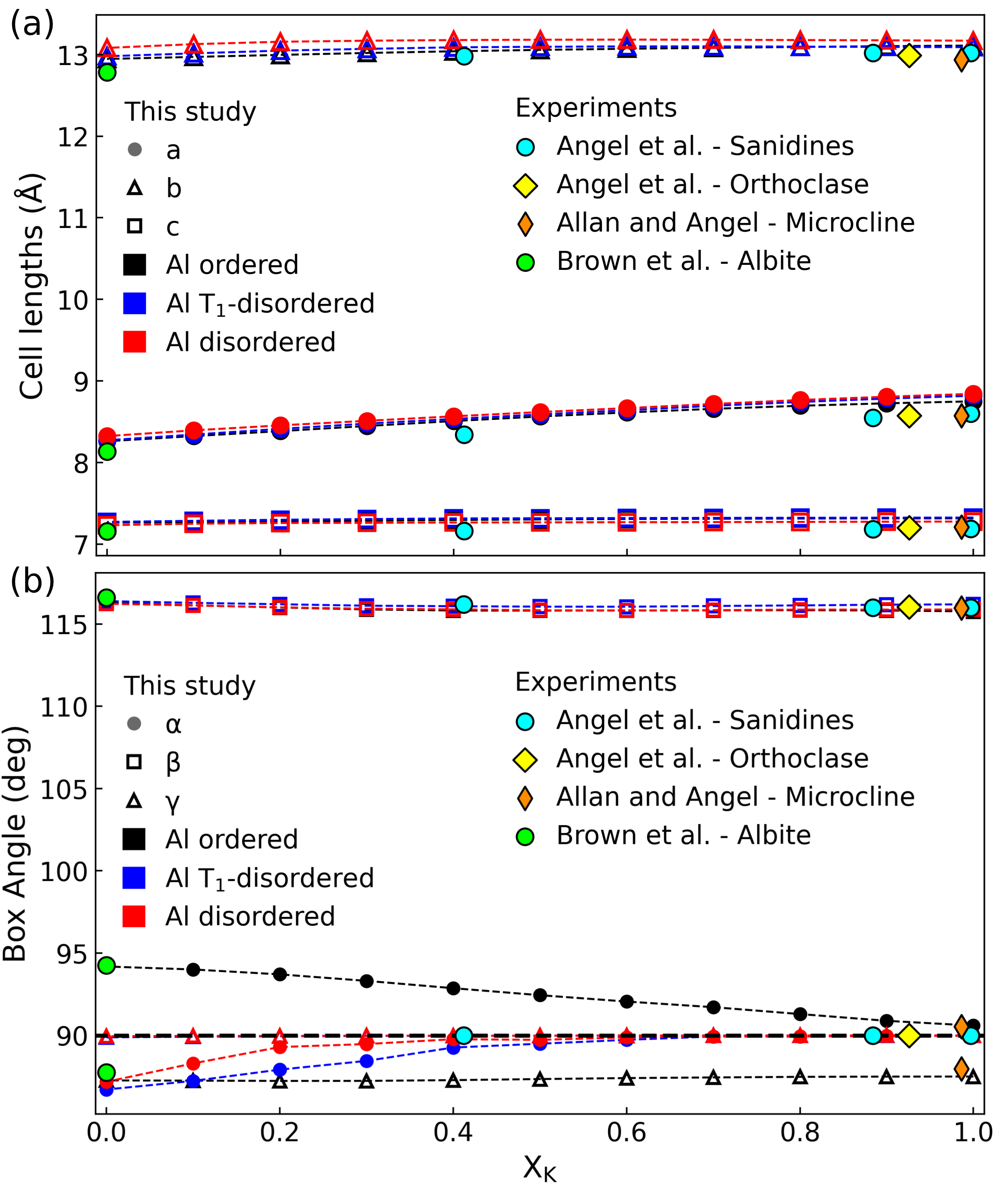
    }}
    \caption{Compositional dependence of the unit cell parameters for the three different states of aluminum order-disorder. With increasing potassium site fraction \(X_{\ch{K}}\),  the solid-solution  becomes successively  more monoclinic. For the fully ordered Al-Si framework, the crystal is triclinic for all \(X_{\ch{K}}\). For both the \(\mathrm{T}_{1}\)-disorder and the fully Al disordered system the angle  \(\alpha \leq \SI{90}{\deg}\) implies triclinic symmetry at \(X_{\ch{K}} \leq 0.3\) transitioning to monoclinic symmetry towards higher  \(X_{\ch{K}}\). We compare with experiments of Angel et al.~\cite{angel_structural_2013}, Allan and Angel~\cite{allan_high-pressure_1997} and Brown et al.~\cite{brown_triclinic_2006}.}  \label{fig:unit_cell_parameters}
\end{figure}

Alkali feldspar transitions between triclinic and monoclinic symmetry, depending on temperature, the state of Al-Si disorder and potassium site fraction \(X_{\ch{K}}\). For monoclinic symmetry, \(\alpha = \gamma = \SI{90}{\deg}\). The compositional dependence of \(\alpha\),  \(\beta\) and  \(\gamma\)  is shown in Fig.~\ref{fig:unit_cell_parameters}~(b). For high and intermediate states of Al-Si disorder the crystal structure is triclinic for Na-rich compositions and approaches monoclinic symmetry with increasing potassium site fraction, a trend that is consistent with experimental observations. The ordered system never reaches monoclinic symmetry even at \(X_{\ch{K}}= 1\), which aligns with the behavior of microcline, fully ordered K-feldspar. The two Al-Si disordered systems are much closer to the monoclinic symmetry overall and are fully monoclinic at  \(X_{\ch{K}} \geq 0.4\). The differences between the Al \(\mathrm{T}_{1}\)-disorder and the full disorder are subordinate.

\subsection{Miscibility and non-ideality}

\begin{figure}
    \centering
    \makebox[\columnwidth][c]{\includegraphics[width=1\columnwidth]{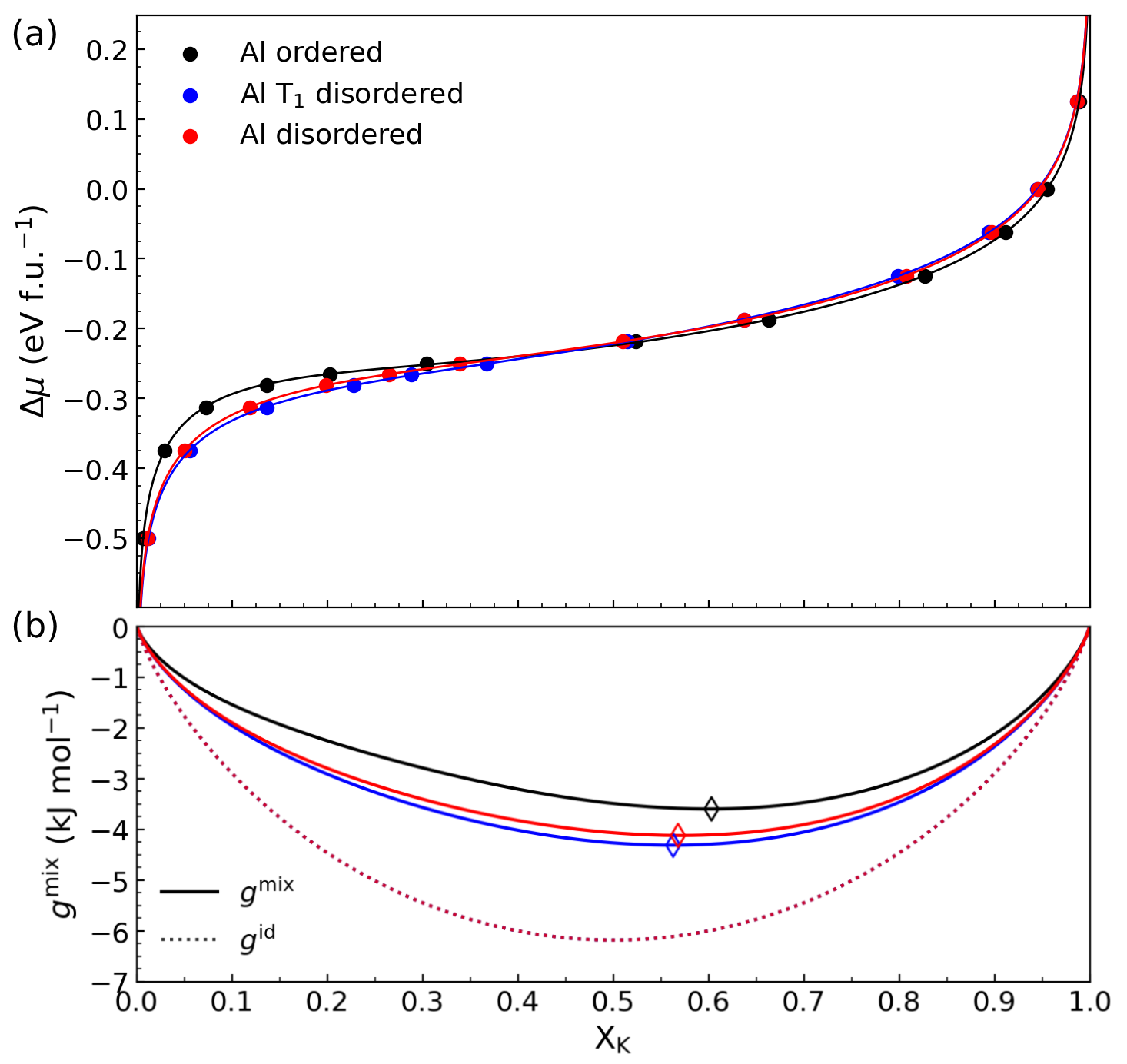
    }}
    \caption{ \(\mathbf{(a)}\) Difference in chemical potential \(\Delta \mu = \mu_{\ch{Na}} - \mu_{\ch{K}}\) versus equilibrium composition \(X_{\ch{K}}\) of alkali feldspar for the three states of Al-Si disorder at \SI{1073.15}{\kelvin}. The definitive integral of \(\Delta \mu (X_{\ch{K}})\) and subtraction of the linear dependence yields the Gibbs energy of mixing \(g^{\mathrm{mix}}\) shown in \(\mathbf{(b)}\). The extrema are shown as diamonds and the Gibbs energy of mixing  of an ideal mixture \(g^{\mathrm{id}}\) is shown as a dotted line.}   \label{fig:mu_C_gmix}
\end{figure}

The results of the SGCMC-MD simulations are shown in Fig.~\ref{fig:mu_C_gmix}. In (a) the relationship between chemical potential difference \(\Delta \mu\) and the equilibrium K concentration \(X_{\ch{K}}\) is presented for the three states of Al-Si disorder. The shapes of the curves indicate that the non-ideality is more pronounced for the Al-Si ordered state than for the two disordered ones. The latter two align closely, with the Al \(T_{1}\)-disordered state being slightly more ideal. The results from SGCMC modeling are %
similar to the equilibrium partitioning experiments between alkali feldspar and a \ch{NaCl}-\ch{KCl} salt melt, which yields the equilibrium \({X_{\ch{K}}}\) site fraction of alkali feldspar as a function of melt composition as expressed by its molar \ch{KCl} fraction~\cite{heuser_thermodynamic_2024}. 

\begin{figure}[h!]
    \centering
    \makebox[\columnwidth][c]{\includegraphics[width=1\columnwidth]{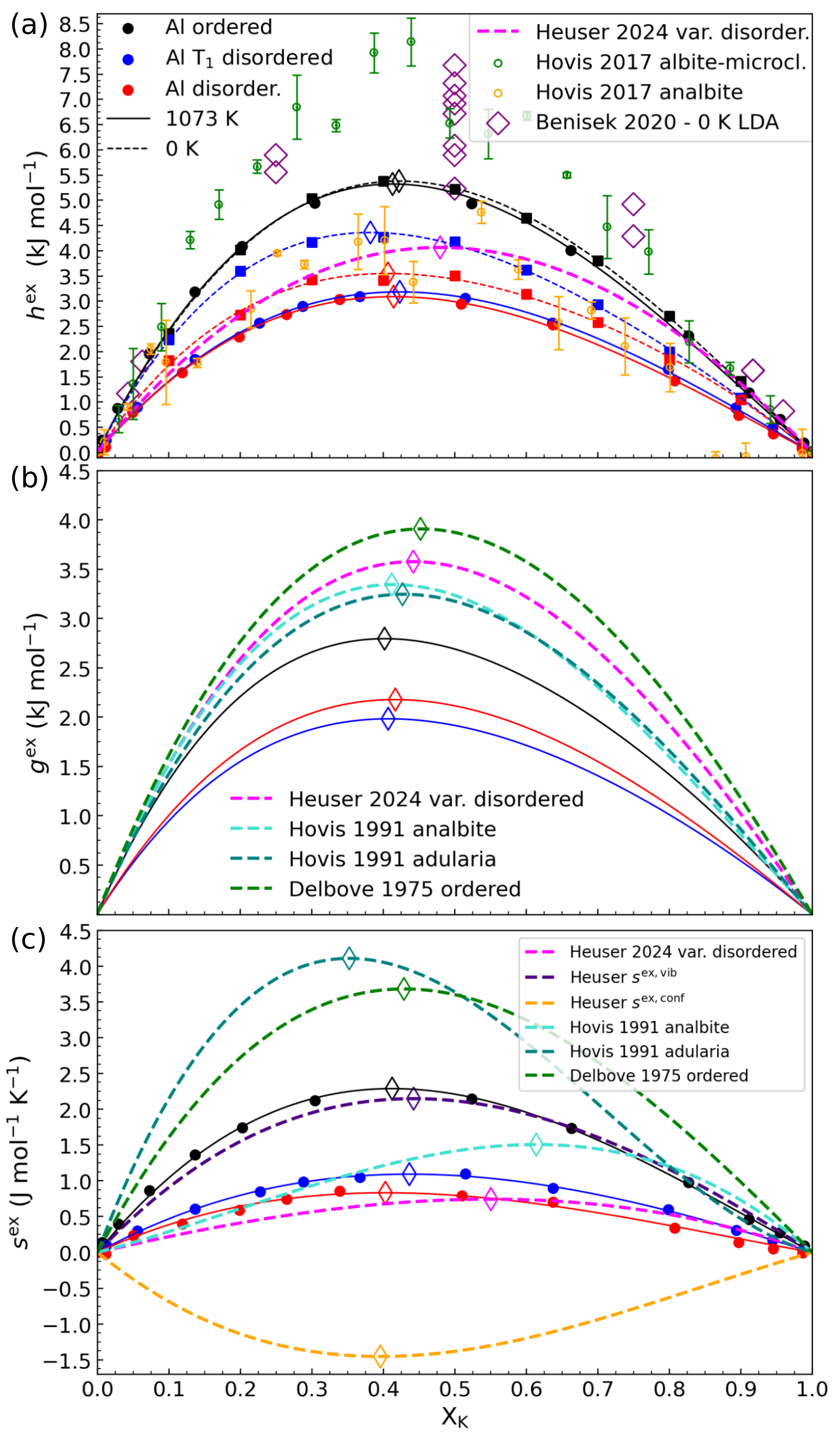
    }}
    \caption{Excess properties of mixing versus composition for the different states of Al-Si disorder. The results of this study are in black/blue/red for the Al-Si ordered/\(\mathrm{T}_1\)-disordered/disordered systems. \(\mathbf{(a)}\) shows the excess enthalpy of mixing \(h^{\mathrm{ex}}\) sampled over the simulations at \SI{1073.15}{\kelvin} and at \SI{0}{\kelvin} (dashed). In \(\mathbf{(b)}\) the excess Gibbs energy of mixing \(g^{\mathrm{ex}}\) obtained from Fig.~\ref{fig:mu_C_gmix}~(b) is shown. \(\mathbf{(c)}\) shows the excess entropy of mixing \(s^{\mathrm{ex}} = (h^{\mathrm{ex}}-g^{\mathrm{ex}} )/T\). Diamonds locate the extrema. Curves in (a) and (c) are the Margules fit to our results.~~~ We compare with experiments of Heuser et al.~\cite{heuser_thermodynamic_2024} (\SI{1073.15}{\kelvin}), Hovis 2017 \cite{hovis_refined_2017} (\SI{323.15}{\kelvin}) for ordered \(h^{\mathrm{ex}}\) (albite-microcline) and disordered \(h^{\mathrm{ex}}\) (analbite), LDA calculations of Benisek and Dachs~\cite{benisek_excess_2020} (\SI{0}{\kelvin}) for ordered \(h^{\mathrm{ex}}\), Hovis et al.~1991~\cite{hovis_gibbs_1991} (\SI{1073.15}{\kelvin}) for disordered (adularia) and medium-disordered (analbite) \(g^{\mathrm{ex}}\) and \(s^{\mathrm{ex}}\) and Delbove~\cite{delbove_excess_1975} (\SI{1073.15}{\kelvin}) for ordered \(g^{\mathrm{ex}}\) and \(s^{\mathrm{ex}}\).}   \label{fig:g_exc}
\end{figure}

Figure \ref{fig:g_exc} shows the excess properties of mixing. The enthalpy of mixing \(h^{\mathrm{ex}}\) in (a) was determined from the potential energies and volumes sampled during the SGCMC-MD simulations at \SI{1073.15}{\kelvin}, as well as from the potential energy of the fully relaxed systems at \SI{0}{\kelvin}. We can see that introducing Al-Si disorder leads to more ideal values as the maximum is shifted almost \SI{2.5}{\kilo\joule\per\mole} lower at \SI{1073.15}{\kelvin} and between  1 to \SI{2}{\kilo\joule\per\mole} lower at \SI{0}{\kelvin}. This observation holds for both Al-Si disordered states, which show remarkable agreement at high temperatures. Possibly, the excess enthalpy decreases when Al distributes on both  \(\mathrm{T}_\mathrm{O}\)- and \(\mathrm{T}_\mathrm{M}\)-tetrahedra, regardless of whether these fall under the \(\mathrm{T}_{1}\) or the \(\mathrm{T}_{2}\) grouping. As temperature increases, the solution becomes more ideal as evidenced by a decrease in \(h^{\mathrm{ex}}\). This effect depends significantly on the state of Al-Si disorder, with the Al-Si ordered state exhibiting almost no and the Al \(\mathrm{T}_1\) disordered state the largest shift. The location of the maximum remains constant at around  \(X_{\ch{K}}=0.41\) for all \(h^{\mathrm{ex}}\) curves.

Let us contextualize these results of \(h^{\mathrm{ex}}\). Our disordered \(h^{\mathrm{ex}}\) calculations agree quite well with the 2017 experiment of Hovis~\cite{hovis_refined_2017} on analbite (disordered albite) over the entire compositional range. Also the \(h^{\mathrm{ex}}\) determined by Heuser and coworkers~\cite{heuser_thermodynamic_2024} for various kinds of Al-disordering using the same conditions (\SI{1073.15}{\kelvin} and 1~atm) is in a similar range to Hovis. For the case of ordered alkali feldspar we can compare with Hovis's albite-microcline. At close to end-member compositions our calculations are in good agreement with the experimental data, but between \(0.2\)-\(0.8\) the experimental \(h^{\mathrm{ex}}\) is significantly underestimated. In a recent study by Benisek and Dachs~\cite{benisek_excess_2020} \(h^{\mathrm{ex}}\) at \SI{0}{\kelvin} was calculated using the same methodology, the only difference being the use of DFT directy, albeit using a different functional; a local-density approximation (LDA) instead of a generalized gradient ascent (GGA) with which we trained our NNP on. The substantially smaller system sizes that are accessible for direct application of DFT result in a large spread and only the Al-ordered system was studied. Nonetheless, the result of Benisek and Dachs seems to be in slightly better agreement with the experiment of Hovis on the albite-microcline \(h^{\mathrm{ex}}\) than our Al ordered ones, but also underestimating the data.
Noteworthy is also that away from the maximum, the data of Heuser and coworkers seems to closely match our disordered \(h^{\mathrm{ex}}\) at low \(X_{\ch{K}}\) and with our ordered \(h^{\mathrm{ex}}\) at high \(X_{\ch{K}}\).

The definitive integral of \(\Delta \mu (X_{\ch{K}})\) yields the Gibbs energy of mixing plotted in Fig.~\ref{fig:mu_C_gmix}~(b). Subtracting the Gibbs energy  of an ideal mixture \(g^{\mathrm{id}}(X_{\ch{K}})\) gives the excess Gibbs energy of mixing \(g^{\mathrm{ex}}(X_{\ch{K}})\), which is shown in Fig.~\ref{fig:g_exc}~(b). The maxima of all three ordering states closely align between \(X_{\ch{K}} = 0.40\) to \num{0.41}. This is well in line with the broader literature~\cite{hovis_refined_2017}, which give a maximum between 0.38 and 0.44. However, our absolute \(g^{\mathrm{ex}}\) values are about \SI{1}{\kilo\joule\per\mole} below the experimental findings. An important feature is the effect of the state of Al-Si disorder on \(g^{\mathrm{ex}}\). The experimental data is inconclusive regarding this. While the experiment of Heuser and coworkers, which included a range  of Al-Si disordered states, found no difference in excess Gibbs energy of mixing, a collection of investigations covering a broader range of states of Al-Si disorder including Al-Si ordered states does suggest an increase of excess Gibbs energy of mixing with increasing Al-Si order. Our data may offer an explanation: \(g^{\mathrm{ex}}\) is markedly higher for the Al-Si ordered case (consistent with Hovis~\cite{hovis_refined_2017}), but the two Al-Si disordered states are almost equivalent (consistent with Heuser et al.~\cite{heuser_thermodynamic_2024}), suggesting that the \(g^{\mathrm{ex}}\) primarily depends on the existence of Al-Si disorder but less so on the specific \(\mathrm{T}_1\)-\(\mathrm{T}_2\) state of this disorder, which is similar to what we found for \(h^{\mathrm{ex}}\).

Finally, the excess entropy of mixing \(s^{\mathrm{ex}} = (h^{\mathrm{ex}} - g^{\mathrm{ex}})/T\) was calculated and is shown in Fig.~\ref{fig:g_exc}~(c). The maxima are located between \(X_{\ch{K}} = 0.40\) to 0.42. For the case of disorder the \(s^{\mathrm{ex}}\) fit Heuser et al.~\cite{heuser_thermodynamic_2024} quite well. In our data the \(s^{\mathrm{ex}}\) increases with increasing order. It is not clear from the experiments, whether this should be expected as \(s^{\mathrm{ex}}\) of the low albite-microcline of Delbove~\cite{delbove_excess_1975} is certainly higher than the \(s^{\mathrm{ex}}\) for the various disordered specimen but the 1991 data of Hovis and coworkers~\cite{hovis_gibbs_1991} on adularia (intermediate disorder) reaches very high \(s^{\mathrm{ex}}\) which contrasts with Heuser~\cite{heuser_thermodynamic_2024}.

\subsection{Short range order}

\begin{figure}
    \centering
    \makebox[\columnwidth][c]{\includegraphics[width=1\columnwidth]{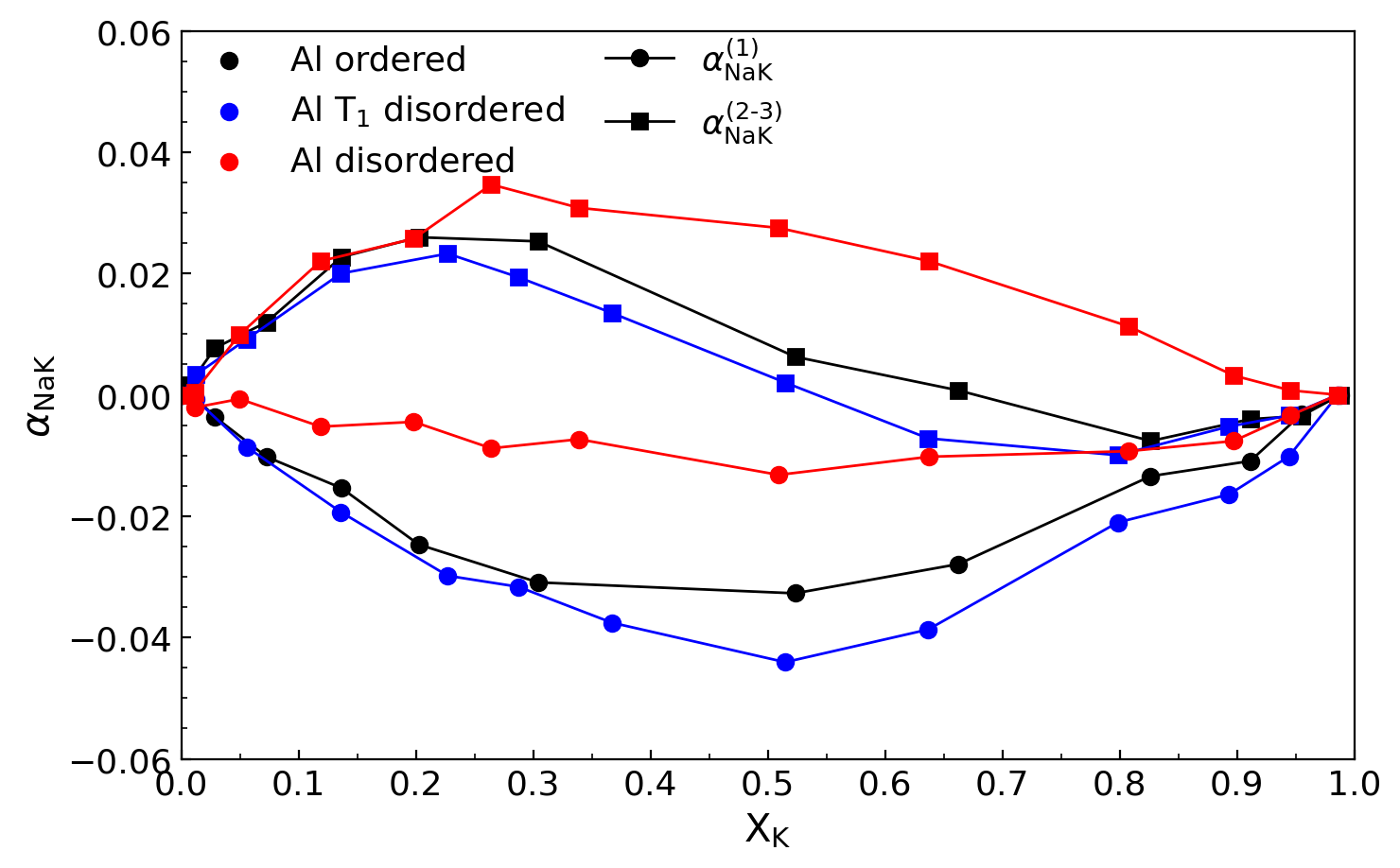
    }}
    \caption{Short range order parameters for the different ordering states.}   \label{fig:sro}
\end{figure}

We calculated the Warren-Cowley SRO parameters \(\alpha_{\ch{NaK}}^{(l)}\) for the first three alkali coordination shells illustrated in Fig.~\ref{fig:uni_cell}~(b) using Eq.~\ref{eq:SRO} and show the result in Fig.~\ref{fig:sro}. %
The second and third coordination shell are treated as if they were one shell due to their similarities.
The shells exhibit opposite coordination tendencies: \(\alpha_{\ch{NaK}}^{(1)} < 0\) indicating fewer \ch{Na}-\ch{Na} nearest neighbors than expected for a random distribution, while \(\alpha_{\ch{NaK}}^{(2-3)} > 0\) indicating a surplus of \ch{Na}-\ch{Na} next nearest neighbors. 

This Na-K arrangement is illustrated in an exaggerated fashion in Fig.~\ref{fig:uni_cell}~(b). We can see that, were the \(\alpha_{\ch{NaK}}^{(1)}\) and \(\alpha_{\ch{NaK}}^{(2-3)}\) at their maximum values, we would see Na-K lamellae parallel to the \((100)\) plane. This orientation is similar (off by around \SI{24}{\deg}) to the perthite lamellar intergrowth that forms along (-801) to (-601), in which also long-range elastic interactions play a  role~\cite{petrishcheva_coherent_2023}, which we do not expect to influence the SRO. We therefore consider the SRO as a precursor to the perthite lamellae. That a certain SRO can be thought of as a precursor (or remnant) of the low-temperature long-range-ordered phase happens frequently in binary alloys~\cite{wolverton_short-range-order_2000} such as \ch{Cu_{1-x}Au_x}~\cite{wolverton_first-principles_1998}.

The SRO parameters, although small, are not negligible, especially considering the high temperature. For comparison, Muzyk et al.~\cite{muzyk_phase_2011} reported minimum \(\alpha_{\ch{WV}}^{(1)}\) and \(\alpha_{\ch{WV}}^{(2)}\) values of around $-0.12$ and $-0.08$, respectively, for tungsten-vanadium alloys at \SI{1000}{\kelvin} and associated these with non-ideal mixing properties. The non-zero SRO parameters in our study similarly imply excess mixing properties in alkali feldspar, even at high temperatures. Our results also indicate a slightly stronger ordering at \(X_{\ch{K}} < 0.5\) as compared to more K-rich compositions, consistent with the excess configurational entropy determined by Heuser et al.~\cite{heuser_thermodynamic_2024} shown in Fig.~\ref{fig:g_exc}~(c).

To give a rough estimate of \(s^{\mathrm{ex,~\mathrm{conf}}}\) from our SRO we assume that the alloy-solution model is applicable to the alkali sublattice. Then, we can follow a method summarized by Swalin~\cite{swalin_thermodynamic_1972} %
and express the excess configurational entropy as
\begin{equation*}
    s^{\mathrm{ex,~\mathrm{conf}}}\!=\!|\alpha| k_{\mathrm{B}} N_{\mathrm{A}}\!\left[X_{\ch{K}} \ln(X_{\ch{K}}) + (1 - X_{\ch{K}})\ln(1 - X_{\ch{K}}) \right],
\end{equation*}
where \(\alpha\) is a Warren-Cowley type SRO parameter. For Al \(\mathrm{T}_1\) disordered at \(X_{\ch{K}} = 0.51\) our \(\alpha = -0.04\) and we obtain \(s^{\mathrm{ex,~\mathrm{conf}}} =  \SI{-0.23}{\joule\per\kelvin\per\mole}\) which is significant in light of a value of \(s^{\mathrm{ex}}\) of less than \SI{1.1}{\joule\per\kelvin\per\mole} at the same Na-K composition (cf. Fig.~\ref{fig:g_exc}~c).

A direct quantitative comparison with Heuser et al.'s result is not possible. This is because they determined \(s^{\mathrm{ex,~\mathrm{conf}}}\) by subtracting the excess vibrational entropy \(s^{\mathrm{ex,~\mathrm{vib}}}\) determined at \SI{298.15}{\kelvin} from the total \(s^{\mathrm{ex}}\). The values of both \(s^{\mathrm{ex,~\mathrm{conf}}}\) and \(s^{\mathrm{ex,~\mathrm{vib}}}\) are expected to shift at high temperatures, with clustering being canonically much more prevalent at low temperatures.

\subsection{Solvus}
\begin{figure}
    \centering
    \makebox[\columnwidth][c]{\includegraphics[width=1\columnwidth]{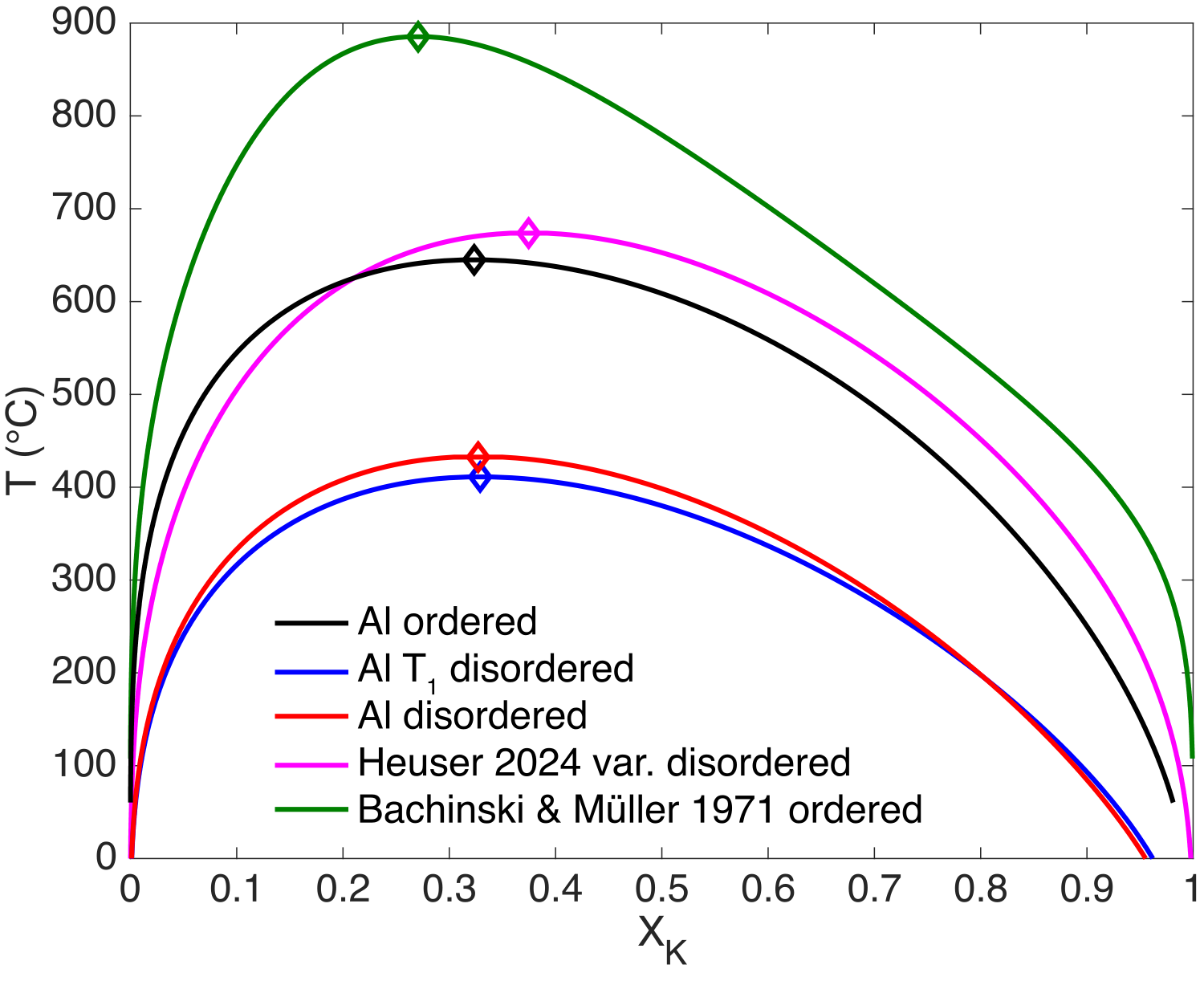
    }}
    \caption{Strain-free solvi of alkali feldspar for the three states of Al-Si disorder and the experiment of Heuser et al.~\cite{heuser_thermodynamic_2024}, in both cases determined at 1~atm as well as the experiment of Bachinski and Müller~\cite{bachinski_experimental_1971} for low albite-microcline determined at \SI{100}{\mega\pascal}.}   \label{fig:solvi}
\end{figure}

Using Margules parameters fitted to the \(h^{\mathrm{ex}}\) and \(s^{\mathrm{ex}}\) of Fig.~\ref{fig:g_exc}~(a,~c) (see Table~\ref{Tab:margules} and the SM), we calculated the strain-free solvus for the three ordering states. The results are shown in Fig.~\ref{fig:solvi}.

All three calculated solvi have a critical point between \(X_{\ch{K}} = 0.32\) to \(0.33\) which aligns well with the experimentally determined solvi of Heuser et al.~\cite{heuser_thermodynamic_2024} at \(X_{\ch{K}} = 0.37\) and of Bachinski and Müller~\cite{bachinski_experimental_1971} at \(X_{\ch{K}} = 0.27\). The critical temperature \(T_\mathrm{C}\) of the ordered solvus is much higher at \SI{645}{\celsius} compared to \SI{411}{\celsius} and \SI{432}{\celsius} for the disordered ones. 

Our ordered solvus is similar to the recently experimentally determined strain-free solvus of Heuser et al.~\cite{heuser_thermodynamic_2024}, which was determined using mixing thermodynamics at \SI{800}{\celsius} to \SI{1000}{\celsius} and 1~atm. Their solvus however represents a range of Al-Si disordered states, and is in agreement with other solvi on disordered alkali feldspars \cite{orville_alkali_1963, luth_alkali_1966, smith_alkali-feldspar_1974} which even though determined at \SI{100}{\mega\pascal}, give a similar result. Bachinski and Müller~\cite{bachinski_experimental_1971} determined the solvus for the Al-Si ordered low albite-microcline series, albeit at \SI{100}{\mega\pascal}, and their \(T_\mathrm{C}\) lies \SI{240}{\celsius} above our ordered solvus. The \(T_\mathrm{C}\) shift between the ordered Bachinski and Müller and disordered Heuser et al.~is marginally smaller at about \SI{211}{\celsius}, which is predicted extremely well in the shift between our ordered and our disordered  \(T_\mathrm{C}\) of \si{212} to \SI{233}{\celsius}. 

In our opinion, the fact that our solvi reproduce the temperature shift between ordered and disordered \(T_\mathrm{C}\) so well, may actually suggest the good agreement of our Al ordered solvus with the solvus of Heuser et al.~\cite{heuser_thermodynamic_2024} to be coincidental. In Fig.~\ref{fig:g_exc}.~(a,~b) we saw that our methodology seems to underestimate the excess mixing properties, but in a way that keeps the effect of Al-Si disordering consistent with the effects observed in experiments. Overall, Figures~\ref{fig:mu_C_gmix} and \ref{fig:g_exc} already suggested that the difference between the Al ordered and disordered states is much bigger than between the various states of disordering. This seems to also be the case for the solvi of Fig.~\ref{fig:solvi}, as a large temperature shift in \(T_\mathrm{C}\) is observed between ordered and disordered in both experiment and simulation, which appears to be less sensitive to variations in the degree of Al-Si disorder as long as disorder prevails. %

\begin{table}
    \centering
    \caption{Margules parameters of the three Al-Si ordering states for the excess enthalpy of mixing  \(W_{h}\) (\si{\kilo\joule\per\mole}) and for the excess entropy of mixing \(W_{s}\) (\si{\joule\per\mole\kelvin}) determined from the \SI{1073.15}{\kelvin} data of Fig.~\ref{fig:g_exc}.}
    \begin{tabular}{l | c c c c}
           & \(W_{h\ch{Na}}\) & \(W_{h\ch{K}}\) & \(W_{s\ch{Na}}\) & \(W_{s\ch{K}}\) \\\toprule
      Al ordered & 12.75 & 28.43 & 5.41 & 12.27 \\
      Al \(\mathrm{T}_1\) disordered  & 8.30  & 16.52 & 3.15 & 5.44 \\
      Al disordered  & 7.48 & 16.44 & 1.79 & 4.59  \\\hline
    \end{tabular}
    \label{Tab:margules}
\end{table}

\section{Conclusions}

The feldspar solid solution poses a formidable challenge to both experiments and simulations. We have demonstrated how effects of both Al\nobreakdash-Si disorder and Na\nobreakdash-K composition can be formally treated starting from first principles. We developed an algorithm that can create alkali-feldspar systems with certain types of Al\nobreakdash-Si disorder which, combined with the construction of a neural network potential, allowed us to probe the impact of Al\nobreakdash-Si disorder on the Na\nobreakdash-K mixing thermodynamics. Excess mixing properties are in good quantitative agreement, and particularly the effects of Al-Si disorder to the excess mixing properties is in great qualitative agreement with past experiments with disorder resulting in a more ideal mixture and the solvus to shift to lower temperatures as compared to Al-Si ordered configurations. Concurrently, the differences between intermediate and full Al disorder seems to be much less pronounced than between ordered and disordered.

Our simulations offer direct evidence for short range ordering of Na\nobreakdash-K which may explain a negative contribution to the configurational entropy measured recently~\cite{hovis_refined_2017, heuser_thermodynamic_2024}. Interestingly, the Warren-Cowley short range order parameter of the first shell \(\alpha_{\ch{NaK}}^{(1)}\) is less than 0 whereas \(\alpha_{\ch{NaK}}^{(2-3)}\) is greater than 0, such that we predict the nearest neighbors to be anticoordinated and the ones further away to be clustered. In an exaggerated fashion, Fig.~\ref{fig:uni_cell}~(b) illustrates that such an ordering goes toward Na-K lamellae parallel to the \((100)\)-plane. Since this is likely as close as SRO can get to the actual perthite lamellae, which are oriented between (-801) and (-601), we consider the short range ordering as a precursor of perthitic lamellar intergrowth.

\section{Acknowledgments}
We would like to thank Lukáš Kývala for suggesting to use \(\chi\) in the wACSF. We acknowledge the financial support from the Austrian Science Fund (FWF) through Grant-DOI 10.55776/I4404 and through Grant-DOI 10.55776/F81. For open access purposes, the author has applied a CC BY public copyright license to any author accepted manuscript version arising from this submission. The computational results presented were achieved using the Vienna Scientific Cluster (VSC).

\section{Data Availability}
The algorithm to create Al-Si and Na-K disordered alkali feldspar structures is available as a Python implementation on the GitHub repository \href{https://github.com/alexgorfer/Alkali-feldspar-disorder-generator}{https://github.com/alexgorfer/Alkali-feldspar-disorder-generator}. The LAMMPS-data files of the relaxed \(8\!\times\!6\!\times\!8\) systems, a template input file for the minimization and for the SGCMC-MD simulation, the training and testing data with and without the point charge correction, the NNP and an n2p2 fork to use the special wACSF to run the NNP are available through the Zenodo repository~\cite{gorfer_thermodynamics_2024}.

\bibliography{library.bib}

\clearpage
\onecolumngrid
\large

\section*{Supplemental Material}
\setcounter{section}{0}
\renewcommand{\thesection}{S-\Roman{section}}
\setcounter{equation}{0}
\renewcommand{\theequation}{S-\arabic{equation}}
\setcounter{table}{0}
\renewcommand{\thetable}{S-\arabic{table}}
\setcounter{figure}{0}
\renewcommand{\thefigure}{S-\arabic{figure}}
\newcounter{SIfig}
\renewcommand{\theSIfig}{S-\arabic{SIfig}}

\vspace{-10pt}
\subsection{System snapshots}
\begin{figure}[h]
    \centering
    \makebox[\columnwidth][c]{\includegraphics[width=1\columnwidth]{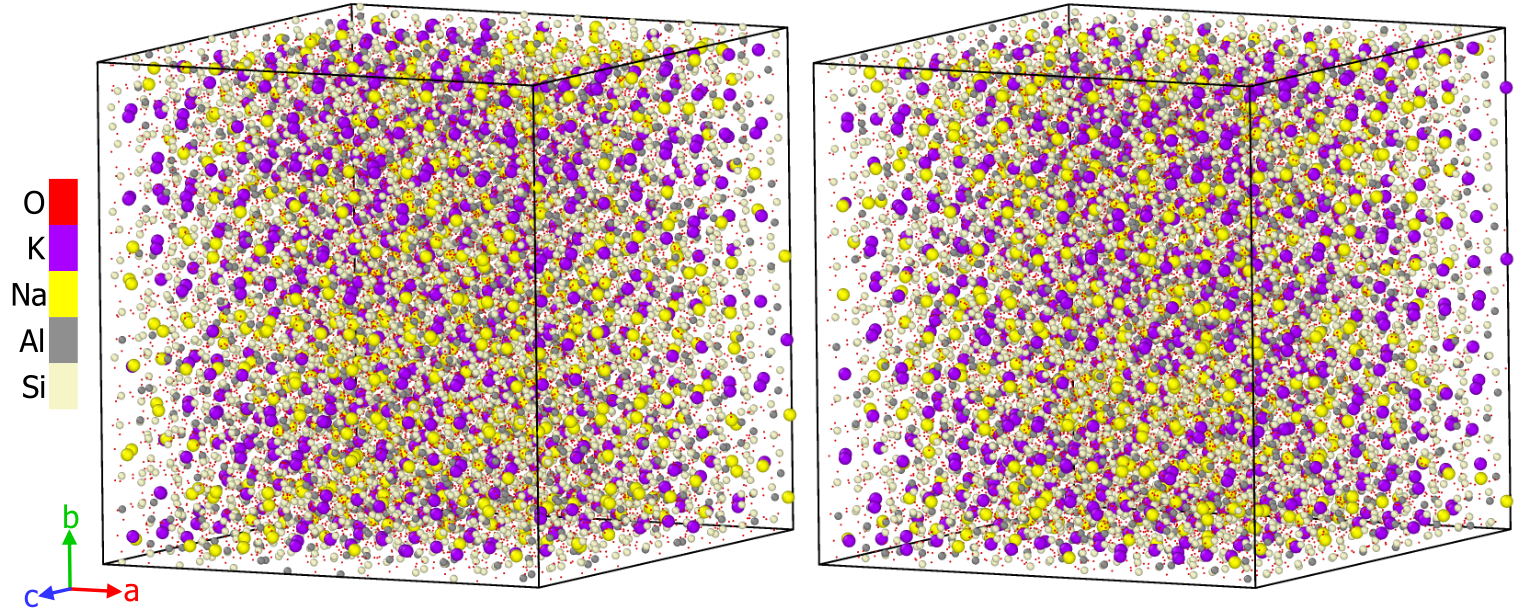
    }}
    \caption{Snapshots of the Al disordered alkali feldspar during the SGCMC-MD at \SI{1073.15}{\kelvin} for \(\Delta \mu = \SI{-0.021875}{\electronvolt}\). The two snapshots are \SI{160}{\pico\second} or \num{100000} attempted and \num{26576} accepted MC moves apart.}   \label{fig:system_SI}
\end{figure}

\subsection{Functional fits}
\subsubsection{Figure 4 and Figure 5 (b)}
The chemical potential versus \(X_{\ch{K}}\), computed using the SGCMC-MD and shown in Figure 4 (a), was fitted with
\begin{equation}\label{eq:deltamu}
    \Delta \mu = k_\mathrm{B} T \log\left[ \frac{X_{\ch{K}}}{1 - X_{\ch{K}}} \right] + \sum_{i = 0}^n A_i  X_{\ch{K}}^i,
\end{equation}
where we used a third order polynomial, i.e.~\(n = 3\). The Gibbs energies of mixing of Figure 4 (b) were then determined by
\begin{equation}
    g^{\mathrm{mix}} = k_\mathrm{B} T \left( X_{\ch{K}} \log X_{\ch{K}} + (1 - X_{\ch{K}}) \log \left[1 - X_{\ch{K}}\right] \right) + \sum_{i=0}^n \frac{A^i X_{\ch{K}}^{i+1}}{i + 1} - X_{\ch{K}} \sum_{i=0}^n \frac{A^i}{i + 1},
\end{equation}
where the first two terms are just the integral of Eq.~\ref{eq:deltamu} and in the last we subtract the linear dependency. If we leave out the first term, the ideal Gibbs free energy without linear dependency, we get the excess Gibbs free energy of mixing:
\begin{equation}\label{Eq:gibbs_ex}
    g^{\mathrm{ex}} = \sum_{i=0}^n \frac{A^i X_{\ch{K}}^{i+1}}{i + 1} - X_{\ch{K}} \sum_{i=0}^n \frac{A^i}{i + 1},
\end{equation}
shown in Fig. 5 (b).
\subsubsection{Figure 5 (a,c) - Margules mixing model}
The excess enthalpy of mixing \(h^{\mathrm{ex}}\) and the excess entropy of mixing \(s^{\mathrm{ex}}\) (but not the \(g^{\mathrm{ex}}\)) were fitted using a Margules mixing model
\begin{equation}
\begin{split}
    h^{\mathrm{ex}} &= X_{\ch{K}} (1 - X_{\ch{K}}) \left[W_{h\ch{K}} (1 - X_{\ch{K}}) + W_{h\ch{Na}}X_{\ch{K}}\right]\\
    s^{\mathrm{ex}} &= X_{\ch{K}} (1 - X_{\ch{K}}) \left[W_{s\ch{K}} (1 - X_{\ch{K}}) + W_{s\ch{Na}}X_{\ch{K}}\right],
\end{split}
\end{equation}
yielding the Margules parameters \(W_{h\ch{K}}, W_{h\ch{Na}}, W_{s\ch{K}}\) and \(W_{s\ch{Na}}\) given in Table~\ref{Tab:margules}. Equation~\ref{Eq:gibbs_ex} is not of the Margules form, but we can use the relation
\begin{equation}
    W_g = W_h - W_s T,
\end{equation}
to also determine \(W_{g\ch{K}}\) and  \(W_{g\ch{Na}}\).
\subsection{Special weighted atom centered symmetry functions}
The Atom Centered Symmetry Functions (ACSF) introduced in \cite{behler_generalized_2007, behler_atom-centered_2011} are the two-body or radial ACSF \(G^2_i\) and the three-body or angular ACSF; narrow angular \(G^3_i\) and wide angular \(G^9_i\):
\begin{equation*}
\begin{split}
    G^2_i &= \sum_{j \neq i} \exp{\left[-\eta (r_{ij} - r_s)^2\right]} f_c(r_{ij})\\
    G^3_i &= 2^{1 - \zeta} \!\!\!\!\! \sum_{j,k \neq i,~j<k} (1 + \lambda \cos \theta_{ijk})^\zeta  \exp{\left[-\eta \left( (r_{ij} - r_s)^2 + (r_{ik} - r_s)^2 + (r_{jk} - r_s)^2 \right)\right]} f_c(r_{ij}) f_c(r_{ik}) f_c(r_{jk}),\\
    G^9_i &= 2^{1 - \zeta} \!\!\!\!\! \sum_{j,k \neq i,~j<k} (1 + \lambda \cos \theta_{ijk})^\zeta \exp{\left[-\eta \left( (r_{ij} - r_s)^2 + (r_{ik} - r_s)^2 \right)\right]} f_c(r_{ij}) f_c(r_{ik}),
\end{split}
\end{equation*}
where \(r_{ij}\) is the distance vector between the central atom \(i\) and its neighbor \(j\), \(\eta\) is a hyperparameter for the shape of the Gaussians, \(\lambda\) centers the cosine terms at the three-body angle \(\theta_{ijk}\) at \SI{0}{\deg} or \SI{180}{\deg}, \(2^{1 - \zeta}\) is a normalization factor, \(r_s\) shifts the center of the Gaussians and \(f_c\) is a smooth cutoff function. We refer to~Ref.~\cite{behler_four_2021} for more details and note that the above equations use the notation of n2p2~\cite{singraber_library-based_2019}.

Chemically diverse systems can pose a challenge when using the angular ACSF as the number of combinations of elements grows quadratically with number of chemical elements such that around five elements the use of ACSF becomes computationally unwieldy. An alternative symmetry function type to amend this issue was introduced by Gastegger et al.~\cite{gastegger_wacsfweighted_2018} in the form of weighted ACSF \(W^3\) in which an additional prefactor \(h(A_j, B_k)\) is introduced inside the sum of \(G^3_i\):

\begin{equation*}
\begin{split}
    W^3_i &= 2^{1 - \zeta} \!\!\!\!\! \sum_{j,k \neq i, j<k} \!\!\!\!\! h(A_j, B_k) (1 + \lambda \cos \theta_{ijk})^\zeta \exp{\left(-\eta \left( (r_{ij} - r_s)^2 + (r_{ik} - r_s)^2 + (r_{jk} - r_s)^2 \right)\right)} f_c(r_{ij}) f_c(r_{ik}) f_c(r_{jk})\\
\end{split}
\end{equation*}

where \(A_j, B_k\) are the atom-types for one pair of neighbors. In \cite{gastegger_wacsfweighted_2018} the function \(h(A_j, B_k)\) was identified with:
\begin{equation*}
    h^{Z}(A_j, B_k) \equiv Z_j Z_k,
\end{equation*}
where \(Z\) is the atomic number. After preliminary tests in using \(h^{Z}\) for the construction of the NNP for alkali feldspar (\ch{O} \ch{Al} \ch{Si} \ch{Na} \ch{K}) we were prompted to search for alternative definitions of \(h(A_j, B_k)\). After multiple trials of different definitions the best performing choice was the use of \(G^3_i\) and \(G^9_i\) for angles involving \ch{O} together with \(W^3_i\) where \(h\) is:
\begin{equation*}
    h^{\chi}(A_j, B_k) \equiv \chi_j \chi_k,
\end{equation*}
where \(\chi\) is the Pauling electronegativity. For example, 
\begin{equation*}
    h^{\chi}(\ch{Na}, \ch{Al}) = 0.92 \cdot 1.61 = 1.48
\end{equation*}

We believe this is a reasonable choice of \(h\) as \(\chi\) is certainly a better signifier of chemical uniqueness than \(Z\) and suspect that this could be a better choice of \(h(A_j, B_k)\) in general.

\subsection{Electrostatic correction for charged defects}
The inclusion of charged defects in the training data warrants the use of corrections to mitigate electrostatic finite size effects. To the potential energies that result from DFT calculations, we therefore added a correction term. This was explored in our previous study~\cite{gorfer_structure_2024}, where we used the Kumagai and Oba correction. That correction is however not feasible to do for the present NNP as one would formally need to determine a different correction for every realization of Al-Si and Na-K disorder. The point charge correction \(E_{\mathrm{PC}}\) was used instead. 

To calculate the correction we need the Madelung potential for a charge \(q\) at \(\mathbf{r} = \mathbf{0}\) for a general anisotropic dielectric tensor \(\overline{\epsilon}\) \cite{kumagai_electrostatics-based_2014}~(Eq.~8):
\begin{equation*}
    \begin{split}
        V_{\mathrm{PC}, q}^\mathrm{aniso} (\mathbf{r} = \mathbf{0}) = &\sum_{\mathbf{R}_i}^{i\neq0} \frac{q}{\sqrt{|\overline{\epsilon}|}} \frac{\mathrm{erfc}\left(\gamma\sqrt{\mathbf{R}_i\cdot \overline{\epsilon} ^{-1} \cdot \mathbf{R}_i }\right)}{\sqrt{\mathbf{R}_i\cdot \overline{\epsilon} ^{-1} \cdot \mathbf{R}_i }} - \frac{\pi q}{\Omega \gamma}\\
            & + \sum_{\mathbf{G}_i}^{i\neq 0} \frac{4 \pi q}{\Omega} \frac{\mathrm{exp}\left(-\mathbf{G}_i\cdot \overline{\epsilon} \cdot \mathbf{G}_i/4\gamma^2 \right)}{\mathbf{G}_i\cdot \overline{\epsilon} \cdot \mathbf{G}_i }  - \frac{2\gamma q}{\sqrt{\pi|\overline{\epsilon}|}},\\
    \end{split}
\end{equation*}
where \(\mathbf{R}_i\) are lattice vectors, \(\mathbf{G}_i\) are reciprocal lattice vectors, \(\mathbf{R}_0\) and \(\mathbf{G}_0\) are \(\mathbf{0}\), \(\gamma\) is a convergence parameter and \(\Omega\) the supercell volume.

Then the point charge correction if given by:
\begin{equation*}
    E_{\mathrm{PC}} = - \frac{q}{2} V_{\mathrm{PC}, q}^\mathrm{aniso} (\mathbf{r} = \mathbf{0}).
\end{equation*}

The charge depends on the inserted defect. For the case of the interstitial defects (Na-Na-dumbbell, \ch{Na}-\((0,0,\frac{1}{2})\)-interstitial, Na-K-dumbbell) \(q\) has a value of \(+1\). For the case of the vacancies (Na-vacancy and K-vacancy) \(q\) has a value of \(-1\). For the case of the defect-free system, no correction is necessary, i.e.~\(q\) has a value of 0.

The dielectric tensor \(\overline{\epsilon}\) is a material-dependent property. We do not consider Al-Si disorder effects on the dielectric tensor. For the case of ordered pure Na-feldspar, the dielectric tensor \(\overline{\epsilon}_{\ch{Na}}\) was calculated from DFT in \cite{gorfer_structure_2024}. Here we repeated the same DFT perturbation routines implemented in VASP~\cite{gajdos_linear_2006} for ordered pure K-feldspar, yielding the \(\overline{\epsilon}_{\ch{K}}\). %

The dielectric tensor for intermediate Na-K compositions \(X_{\ch{K}}\) was then approximated using the interpolation
\begin{equation*}
    \overline{\epsilon}(X_{\ch{K}}) = (1 - X_{\ch{K}}) \overline{\epsilon}_{\ch{Na}} + X_{\ch{K}}\overline{\epsilon}_{\ch{K}}.
\end{equation*}

\end{document}